\documentclass[sort&compress]{elsarticle}
\usepackage[english]{babel}
\usepackage{graphicx}
\usepackage{makeidx}
\usepackage{amsmath,natbib}
\usepackage{breqn}
\usepackage{mathptmx}

\journal{arXiv}

\begin{document}

\begin{frontmatter}

\title{The neutron-gamma Feynman variance to mean approach: gamma detection and total neutron-gamma detection (theory and practice)}

\author[rvt]{Dina Chernikova\corref{cor1}}
\author[rvt,focal]{K{\aa}re Axell}
\author[td]{Senada Avdic}
\author[rvt]{Imre P\'{a}zsit}
\author[rvt]{Anders Nordlund}
\author[rvtd]{Stefan Allard}

\cortext[cor1]{Corresponding author, email: dina@nephy.chalmers.se}
\address[rvt]{Chalmers University of Technology, Department of Applied Physics, Nuclear Engineering, \\Fysikg{\aa}rden 4, SE-412 96 G\"oteborg, Sweden}
\address[focal]{Swedish Radiation Safety Authority, SE-171 16 Stockholm, Sweden}
\address[td]{University of Tuzla, Department of Physics, 75000 Tuzla, Bosnia and Herzegovina}
\address[rvtd]{Chalmers University of Technology, Department of Chemical and Biological Engineering, Nuclear Chemistry and Industrial Materials Recycling, Kemivagen 4, SE-412 96 G\"oteborg, Sweden}

\begin{abstract}

Two versions of the neutron-gamma variance to mean (Feynman-alpha method or Feynman-Y function) formula for either gamma detection only or total neutron-gamma detection, respectively, are derived and compared in this paper. The new formulas have particular importance for detectors of either gamma photons or detectors sensitive to both neutron and gamma radiation. If applied to a plastic or liquid scintillation detector, the total neutron-gamma detection Feynman-Y expression corresponds to a situation where no discrimination is made between neutrons and gamma particles. The gamma variance to mean formulas are useful when a detector of only gamma radiation is used or when working with a combined neutron-gamma detector at high count rates. The theoretical derivation is based on the Chapman-Kolmogorov equation with the inclusion of general reactions and corresponding intensities for neutrons and gammas, but with the inclusion of prompt reactions only. A one energy group approximation is considered. The comparison of the two different theories is made by using reaction intensities obtained in MCNPX simulations with a simplified geometry for two scintillation detectors and a $^{252}$Cf-source. In addition, the variance to mean ratios, neutron, gamma and total neutron-gamma, are evaluated experimentally for a weak $^{252}$Cf neutron-gamma source, a $^{137}$Cs random gamma source and a $^{22}$Na correlated gamma source. Due to the focus being on the possibility of using neutron-gamma variance to mean theories for both reactor and safeguards applications, we limited the present study to the general analytical expressions for Feynman-alpha formulas.

\end{abstract}


\end{frontmatter}

\section{Introduction}

The Feynman-alpha theory for neutrons has a long history in a field of reactor applications. It was traditionally used for evaluation of subcritical reactivity in reactors during startup with a neutron source, and later in accelerator-driven systems \cite{Freya}. Relatively recently this method received attention in the area of nuclear safeguards for purposes of detection of special nuclear materials and spent fuel assay \cite{Twomey,Verbeke1,croft12}. One of the drawbacks of the traditional approach in safeguards applications was related to the use of thermal neutron detectors, which limited the performance in terms of time, type and activity of the radioactive or nuclear material and data completeness (information about the energy of particles is lost during slowing down process). Therefore, attempts were made in order to use fast neutron detection for the evaluation of the Feynman-Y function of the variance to mean ratio \cite{McConchie,Verbeke}. A comparision between the performance of thermal and fast detection systems \cite{Nakae} showed that it took 8.5 hours for a thermal detection system ($^{3}$He counters) in contrast to 100 seconds for a fast detection system (liquid scintillation detectors with pulse shape discrimination) to assay a weak $^{252}$Cf-source by using Feynman-alpha approach. However, when using fast scintillation detectors for this purpose, which are also sensitive to gamma photons, the need arose to discriminate between neutrons and gamma particles, whereby the lack of a robust discrimination method represented a complication in the application of the method.

To address this problem and to extend the capacities of the Feynman-alpha approach, we suggest here an alternative path, by deriving Feynman-alpha formulas by taking into account the generation and detection of gamma photons. This way, two new versions of the Feynman-alpha theory can be elaborated: one based only on the detection of gamma particles, and another in which both neutrons and gamma photons are detected without identifying which is which, i.e. not performing pulse shape discrimination and only considering the total (neutron and gamma) counts. 

In this work we do not include delayed neutrons treatment. The reason for this is that the delayed neutrons are emitted long after the fission event, e.g. mean delay times are varying between 0.1 and 10 s, and therefore, their contribution is expected to be negligible while working with fast detection systems (time scale from 1 ns to 10000 ns). The capture gammas, photofission and photonuclear neutrons are not included in the present theory either. Their effect to the neutron-gamma Feynman variance-to-mean ratios for neutron, gamma and total detection is investigated by us separately \cite{MC2015}. There are two reasons why the new versions of the Feynman-alpha formulas are considered in a one-group approximation. One is to enable them for a relatively easy practical use and another is related to the fact that according to the results of recent theoretical investigation \cite{Dinanew}, the one-point one-group version and the two-point two-group version of Feynman-alpha formulas used for fast neutron detections lead to the same results. 

Having derived the corresponding formulas, they are investigated both numerically and experimentally.

\section{The theoretical background for the neutron-gamma Feynman-alpha theory}

The derivation given below will use the same assumptions as the traditional Feynman-alpha theory for neutrons \cite{Imre}. This approach assumes a homogeneous infinite medium in which the detector is also distrubuted homogeneously. In such a setting the temporal evolution of the neutron chains is determined by the properties of the multiplying medium, which are only very slightly influenced by the presence of the detector, which is a weak absorber compared to the medium. The exponent alpha, extracted from the measurement, is characteristic to the speed of the dying out of the subcritical chains and hence is a characteristics of the material. Due to this reason, the primary use of the theory as developed in this paper is envisaged for enhancing the performance of the Feynman-alpha method for measurements of the subcritical reactivity and it will be tested in measurements at a research reactor.

However, our main purpose is to extend the capabilities of the Feynman-alpha method in safeguards applications in order to characterise an unknown sample. As is also discussed in the literature \cite{croft12}, the application of the Feynman-alpha method in safeguards measurements, and in particular for measurements on orphan sources, is not straightforward. The detector is outside of the (usually small) sample, and only detects the neutrons which leaked out from the sample. In contrast to the large, close to critical reactor cores with a detector embedded into the core or the reflector, the statistics of the leaked out neutrons here is determined largely (and for small samples, solely) by the multiplicity of the neutron source (which is spontaneous fission) and to a lesser extent from the multiplication in the medium, whereas the parameter alpha (inverse of the die-away time) is determined by the properties of the detector. The properties of the material (the unknown sample) are extracted mostly from its properties as a spontaneous neutron source with a multiplicity, rather than from the properties of the internal multiplication in the system.

For these reasons, application of the Feynman-alpha method in safeguards measurements is based on the asymptotic value of the variance to mean, rather than on its temporal evolution, characterised by the parameter alpha. The combination of the asymptotic value of the variance to mean in neutron measurements and the average count rate, was already used for determinining the strength of an$^{252}Cf$ source. One application of the present theory is to attempt the same by using gamma photons or both neutrons and gamma photons. However, we shall have in mind the difference between the assumptions of the model and the application.

\subsection{The main concept and assumptions}

The neutron-gamma variance to mean (Feynman-alpha) formulas for separate gamma detection and total neutron-gamma detection will be derived by using the Kolmogorov forward approach the same way as described in \cite{Imre,Dinanew}. In the model that will be used for the derivations we assume that there are neutron and gamma populations: neutrons (denoted as particles of type 1) and gammas (denoted as type 2). Neutrons can undergo the reactions (i) listed below:
\begin{itemize}
\item absorption (${i=a}$) with no gammas emitted,
\item fission (${i=f}$) with corresponding gamma emission,
\item detection (${i=d}$).
\end{itemize}
Gamma particles can undergo the same reactions (i) as neutrons except fission, i.e. photofission or photonuclear reactions are not considered in the present model.

\begin{figure}[ht!]
\centering
\includegraphics[width=0.45\textwidth]{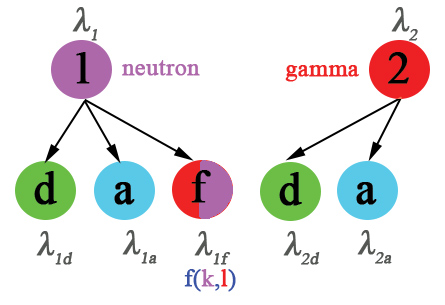}
\caption{A model of various processes which particles can undergo.}
\label{fig:1}
\end{figure}

As mentioned, the assumption behind the model is the same as with the traditional Feynman-alpha theory, i.e. that the medium is infinite and homogeneous with space-independent reaction intensities for absorption of neutrons (${\lambda_{1a}}$) and gammas (${\lambda_{2a}}$), fission induced by neutrons (${\lambda_{1f}}$) and detection of neutrons (${\lambda_{1d}}$) and gammas (${\lambda_{2d}}$), as shown in Figure \ref{fig:1}.

Thus, the total transition intensities for neutrons and gammas are denoted as $\lambda_{1}$ and $\lambda_{2}$:
\begin{equation}
\begin{split}
\lambda_1 = \lambda_{1a} + \lambda_{1f} + \lambda_{1d}\\
\lambda_2 = \lambda_{2a} + \lambda_{2d}
\end{split}
\end{equation}

In the model we include a compound Poisson source of neutrons and gammas with emission intensity $S$. The source is assumed to release $m$ neutrons and $n$ gammas in one emission event with the probability distribution $p (m,n)$. For the induced fission reaction, we consider that $k$ neutrons and $l$ gammas are emitted with the probability distribution $f(k,l)$. All fission neutrons are assumed to be prompt, i.e. delayed neutrons as not considered in this work. No delayed generation processes for gamma photons are assumed either.

\subsection{The one-group one-point model for separate detection of neutrons and gammas}

In order to derive the one-group one-point Feynman-alpha theory for separate detection of neutrons and gammas, let us assume that the source ${S}$ is switched on at the time ${t_{0} \leq t}$, whereas the separate detection processes for neutrons and gammas both are started at the same fixed time instant ${t_{d}}$, where ${t_{0} \leq t_{d} \leq t}$. Let the random processes $N_{1}(t)$ and $N_{2}(t)$ represent the number of neutrons and gammas at the time ${t \geq 0}$, and ${Z_{1}(t,t_{d})}$ and ${Z_{2}(t,t_{d})}$ - the number of neutron and gamma detections in the time interval [${t_{d},t}$], respectively. Thus, the joint probability of having $N_{1}$ neutrons and $N_{2}$ gammas at time $t$ in the system, and having detected  $Z_{1}$ neutrons and $Z_{2}$ gammas during the period of time ${t-t_{d}\geq 0}$ can be defined as $P(N_{1},N_{2},Z_{1},Z_{2},t|t_0)$. By summing up the probabilities of all mutually exclusive events of the particle not having or having a specific reaction within the infinitesimally small time interval d$t$, one obtains the forward Kolmogorov or forward master equation
\begin{equation}
\begin{split}
\frac{dP(N_{1},N_{2},Z_{1},Z_{2},t)}{dt}\\
&=-(\lambda_{1}N_{1}+\lambda_{2}N_{2}+S)P(N_{1},N_{2},Z_{1},Z_{2},t)\\
&+\lambda_{1a}(N_{1}+1)P(N_{1}+1,N_{2},Z_{1},Z_{2},t)+\lambda_{2a}(N_{2}+1)P(N_{1},N_{2}+1,Z_{1},Z_{2},t)\\
&+\lambda_{1f}\sum_k^{N_{1}+1}\sum_l^{N_{2}}(N_{1}+1-k)f(k,l)P(N_{1}+1-k,N_{2}-l,Z_{1},Z_{2},t)\\
&+\lambda_{1d}(N_{1}+1)P(N_{1}+1,N_{2},Z_{1}-1,Z_{2},t)\\
&+\lambda_{2d}(N_{2}+1)P(N_{1},N_{2}+1,Z_{1},Z_{2}-1,t)\\
&+S\sum_m^{N_{1}}\sum_n^{N_{2}}p(m,n)P(N_{1}-m,N_{2}-n,Z_{1},Z_{2},t)
\end{split}
\end{equation}
where, as mentioned earlier, ${f(k,l)}$ is the number distribution of neutrons and gammas in an induced fission event. The initial condition reads as
\begin{equation}
\begin{split}
P(N_{1},N_{2},Z_{1},Z_{2},t=t_0\mid t_0) = \delta_{N_{1},0} \delta_{N_{2},0} \delta_{Z_{1},0} \delta_{Z_{2},0}
\end{split}
\end{equation}
The various moments of the particle numbers and detection numbers can be obtained from this equation by using the generating function technique in a way similar to as described in \cite{Imre}.
By defining the following generating function for the probability distribution $P(N_{1},N_{2},Z_{1},Z_{2},t)$:
\begin{equation}
\begin{split}
G(X,Y,L,W,t)= \sum_{N_{1}} \sum_{N_{2}} \sum_{Z_{1}} \sum_{Z_{2}} X^{N_{1}} Y^{N_{2}} L^{Z_{1}} W^{Z_{2}}P(N_{1},N_{2},Z_{1},Z_{2},t)
\end{split}
\end{equation}
with the initial condition for ${t_0 \leq t}$
\begin{equation}
\begin{split}
G(X,Y,L,W,t=t_0\mid t_0) = 1,
\end{split}
\end{equation}
the following partial differential equation is obtained:
\begin{equation}
\begin{split}
\frac{\partial G}{\partial t}= [\lambda_{1a} +\lambda_{1d}L - \lambda_{1}X + q(X,Y) \lambda_{1f}]\frac{\partial G}{\partial X} + [\lambda_{2a} +\lambda_{2d}W - \lambda_{2}Y]\frac{\partial G}{\partial Y} + S [r(X,Y) - 1] G.
\end{split}
\end{equation}
Here the generating functions of the number distributions of neutrons and gamma photons in a source event (spontaneous fission) and an induced fission event were introduced as
\begin{equation}
\begin{split}
r(X,Y) = \sum_m \sum_n X^m Y^n p(m,n)\\
q(X,Y) = \sum_k \sum_l X^k Y^l f(k,l)
\end{split}
\end{equation}
In the above, as mentioned before, ${p}$\emph{(m,n)} is the probability of having \emph{m} neutrons and \emph{n} gammas produced in a source event. For the sake of simplicity, some notations are introduced as follows:
\begin{equation}
\begin{split}
\left.\frac{\partial}{\partial X}q(X,Y)\right|_{X=Y=1}= \sum_k \sum_l k \cdot f(k,l) =q^{(1,0)}\\
\left.\frac{\partial}{\partial Y}q(X,Y)\right|_{X=Y=1}= \sum_k \sum_l l \cdot f(k,l) =q^{(0,1)}\\
\left.\frac{\partial^2}{\partial X \partial X}q(X,Y)\right|_{X=Y=1}= \sum_k \sum_l k \cdot (k-1) \cdot f(k,l) =q^{(2,0)}\\
\left.\frac{\partial^2}{\partial Y \partial Y}q(X,Y)\right|_{X=Y=1}= \sum_k \sum_l l \cdot (l-1) \cdot f(k,l) =q^{(0,2)}\\
\left.\frac{\partial^2}{\partial X \partial Y}q(X,Y)\right|_{X=Y=1}= \sum_k \sum_l k \cdot l \cdot f(k,l) =q^{(1,1)}
\end{split}
\end{equation}
and
\begin{equation}
\begin{split}
\left.\frac{\partial}{\partial X}r(X,Y)\right|_{X=Y=1}= \sum_m \sum_n m \cdot p(m,n) =r^{(1,0)}\\
\left.\frac{\partial}{\partial Y}r(X,Y)\right|_{X=Y=1}= \sum_m \sum_n n \cdot p(m,n) =r^{(0,1)}\\
\left.\frac{\partial^2}{\partial X \partial X}r(X,Y)\right|_{X=Y=1}= \sum_m \sum_n m \cdot (m-1) \cdot p(m,n) =r^{(2,0)}\\
\left.\frac{\partial^2}{\partial Y \partial Y}r(X,Y)\right|_{X=Y=1}= \sum_m \sum_n n \cdot (n-1) \cdot p(m,n) =r^{(0,2)}\\
\left.\frac{\partial^2}{\partial X \partial Y}r(X,Y)\right|_{X=Y=1}= \sum_m \sum_n m \cdot n \cdot p(m,n) =r^{(1,1)}
\end{split}
\end{equation}
In a steady subcritical medium with a steady source, a stationary state of the system exists when $t_0 \to -\infty$. For that case the following solutions  are obtained for the constant neutron and gamma populations ${\bar{N}_{1}}$, ${\bar{N}_{2}}$ and the time-varying detection counts ${\bar{Z}_{1}(t)}$, ${\bar{Z}_{2}(t)}$:
\begin{equation}\label{eq:10}
\begin{split}
\bar{N}_{1} = \frac{S r^{(1,0)}}{\lambda _1-\lambda _{1f} q^{(1,0)}} \\
\bar{N}_{2} = \frac{S r^{(0,1)}}{\lambda _2}+\frac{S\lambda _{1f} q^{(0,1)}r^{(1,0)}}{\lambda _2(\lambda _1-\lambda _{1f} q^{(1,0)})} \\
\bar{Z}_{1}(t)=\lambda_{1d} \bar{N}_{1} t\\
\bar{Z}_{2}(t)=\lambda_{2d} \bar{N}_{2} t\\
\end{split}
\end{equation}
As a sideline, it can be mentioned that in reactor physics terminology, $\lambda_1 = \lambda_a$ and $\lambda _{1f} q^{(1,0)} = \nu \lambda_f = 1/\Lambda$. Accounting also for the definition of the reactivity $\rho$ as
\begin{equation}
\rho = \frac{\nu \lambda_f - \lambda_a}{\nu \lambda_f},
\end{equation}
the first equation in (\ref{eq:10}) can be written as
\begin{equation}\label{eq:classic}
\bar{N}_{1} = \frac{S \Lambda\,r^{(1,0)}}{-\rho} = \frac{S \Lambda\,r_1}{-\rho},
\end{equation}
where the notation $r_1$ was introduced for the average number neutrons emitted in a source event. Eq. (\ref{eq:classic}) is the known classical result for a stationary subcritical system with a source having multiplicity $r_1$ \cite{Imre,ads1}. This is expected since the coupling between the neutrons and the gamma photons is only one way, i.e. the evolution of the number of gamma photons has no influence on the neutrons, hence all results that refer only to the neutrons are the same as in the classical theory including only neutrons. 

By introducing the modified second factorial moment of the random variables $a$ and $b$ as follows
\begin{equation}
\begin{split}
{\mu_{aa}\equiv\langle a(a-1)\rangle -\langle a\rangle ^{2}}={\sigma_{a}^{2}} - {\langle a\rangle }\\
{\mu_{ab}\equiv\langle ab\rangle -\langle a\rangle \langle b\rangle }
\end{split}
\end{equation}
and then taking cross- and auto-derivatives, the following system of differential equations is obtained for the modified second factorial moments ${\mu _{{{N}_{1}{N}_{1}}}}$, ${\mu _{{{N}_{1}{N}_{2}}}}$, ${\mu _{{{N}_{2}{N}_{2}}}}$ of the neutron and gamma populations:
\begin{equation}\label{eq:stac}
\begin{split}
\frac{\partial}{\partial t} \mu_{{N}_{1}{N}_{1}} = 2(\lambda _{1f} q^{(1,0)}-\lambda _1) \mu_{{N}_{1}{N}_{1}}+\lambda _{1f} q^{(2,0)} \bar{N}_{1}+S r^{(2,0)} \\
\frac{\partial}{\partial t} \mu_{{N}_{1}{N}_{2}} = S r^{(1,1)}+\lambda _{1f} q^{(1,1)} \bar{N}_{1}-\lambda _{2}\mu_{{N}_{1}{N}_{2}} +(\lambda _{1f} q^{(1,0)}-\lambda _1) \mu_{{N}_{1}{N}_{2}}+\lambda _{1f} q^{(0,1)}\mu_{{N}_{2}{N}_{2}}\\
\frac{\partial}{\partial t} \mu_{{N}_{2}{N}_{2}} = S r^{(0,2)}-2\lambda _{2}\mu_{{N}_{2}{N}_{2}} +\lambda _{1f} q^{(0,2)}\bar{N}_{1}+2\lambda _{1f} q^{(0,1)}\mu_{{N}_{1}{N}_{2}}
\end{split}
\end{equation}
In a stationary system, these modified moments are constant, and can be easily obtained by solving the algebraic equation resulting from setting the l.h.s. of (\ref{eq:stac}) equal to zero. The time dependent modified second moments, ${\mu _{{{N}_{1}{Z}_{1}}}}$, ${\mu _{{{N}_{2}{Z}_{1}}}}$, ${\mu _{{{Z}_{1}{Z}_{1}}}}$, ${\mu _{{{N}_{1}{Z}_{2}}}}$, ${\mu _{{{N}_{2}{Z}_{2}}}}$, ${\mu _{{{Z}_{2}{Z}_{2}}}}$ can be found by solving the system of equations:
\begin{equation}
\begin{split}
\frac{\partial}{\partial t} \mu_{{N}_{1}{Z}_{1}} = (\lambda _{1f} q^{(1,0)}-\lambda _1) \mu_{{N}_{1}{Z}_{1}}+\lambda _{1d} \mu _{{N}_{1}{N}_{1}}\\
\frac{\partial}{\partial t} \mu_{{Z}_{1}{Z}_{1}} = 2 \lambda _{1d}\mu_{{N}_{1}{Z}_{1}} \\
\frac{\partial}{\partial t} \mu_{{N}_{2}{Z}_{1}} =-\lambda _2 \mu_{{N}_{2}{Z}_{1}}+\lambda _{1f} q^{(0,1)} \mu_{{N}_{1}{Z}_{1}}+\lambda _{1d} \mu_{{N}_{1}{N}_{2}}\\
\frac{\partial}{\partial t} \mu_{{N}_{1}{Z}_{2}} = (\lambda _{1f} q^{(1,0)}-\lambda _1) \mu_{{N}_{1}{Z}_{2}}+\lambda _{2d} \mu_{{N}_{1}{N}_{2}} \\
\frac{\partial}{\partial t} \mu_{{N}_{2}{Z}_{2}} = -\lambda _2 \mu_{{N}_{2}{Z}_{2}}+\lambda _{1f} q^{(0,1)} \mu_{{N}_{1}{Z}_{2}}+\lambda _{2d} \mu _{{N}_{2}{N}_{2}}\\
\frac{\partial}{\partial t} \mu_{{Z}_{2}{Z}_{2}} = 2 \lambda _{2d}\mu_{{N}_{2}{Z}_{2}}
\end{split}
\end{equation}
Again, as could be expected, the first two equations, referring to neutron numbers and detections only, decouple from the rest, and even from each other such that the first one can be solved alone and the result can be used to solve the second equation. This will lead to the well-known Feynman-alpha expression for neutrons with one exponent only, being equal to $\omega_{n} = \rho/\Lambda$ \cite{ads1}:
\begin{equation}\label{neutronY}
\begin{split}
\frac{\sigma_{Z_1 Z_1}^2(t)}{\bar{Z}_{1}} = 1 + (\frac{\lambda_d (\lambda_1 r^{(2,0)}+\lambda_{1f} (q^{(2,0)}r^{(1,0)}-q^{(1,0)}r^{(2,0)}))}{r^{(1,0)}(\lambda_1-\lambda _{1f} q^{(1,0)})^2}) (1 - \frac{1-e^{- \omega_{n} t}}{\omega_{n} t}) 
\end{split}
\end{equation}

For the gammas, the rest of the coupled equations have to be solved with a second order characteristic equation. The final expression for the Feynman-alpha formulas for gammas is given as below:
\begin{equation}\label{gammaY}
\begin{split}
\frac{\sigma_{Z_2 Z_2}^2(t)}{\bar{Z}_{2}} = 1 + Y_{g1} (1 - \frac{1-e^{- \omega_{g1} t}}{\omega_{g1} t}) + Y_{g2} (1 - \frac{1-e^{- \omega_{g2} t}}{\omega_{g2} t})
\end{split}
\end{equation}
The two roots ${\omega_{g1}}$ and ${\omega_{g2}}$ are obtained as
\begin{equation}
\begin{split}
\omega_{g1}=-\lambda _{1f} q^{(1,0)}+\lambda _1\\
\omega_{g2}=\lambda _2
\end{split}
\end{equation}
It is interesting to notice that ${\omega_{g1}}$ is the same for neutrons and gammas, i.e. $\omega_{g1}=\omega_{n}$.

The functions ${Y_{g1}}$, ${Y_{g2}}$ in the gamma Feynman-alpha formula (\ref{gammaY}) are given in the form:
\begin{equation}
\begin{split}
-Y_{g1}=\frac{2 \lambda _{2d}(\lambda _{1f} \mu _{{N}_{1}{N}_{2}} q^{(0,1)}-\mu _{{N}_{2}{N}_{2}} (\lambda _{1f} q^{(1,0)}+\omega _{g1})+\lambda _1 \mu _{{N}_{2}{N}_{2}})} {\omega _{g1} (\omega _{g1}-\omega _{g2}) \bar{N}_{2}} \\
-Y_{g2}=\frac{2 \lambda _{2d}(\lambda _{1f} \mu _{{N}_{1}{N}_{2}} q^{(0,1)}-\mu _{{N}_{2}{N}_{2}} (\lambda _{1f} q^{(1,0)}+\omega _{g2})+\lambda _1 \mu _{{N}_{2}{N}_{2}})} {\omega _{g2} (\omega _{g2}-\omega _{g1}) \bar{N}_{2}}
\end{split}
\end{equation}
It can be shown that:
\begin{equation}
\begin{split}
Y_{g0} = Y_{g1}+Y_{g2}=\frac{2 \lambda _{2d} (\lambda _{1f} \mu _{{N}_{1}{N}_{2}} q^{(0,1)}-\lambda _{1f} \mu _{{N}_{2}{N}_{2}} q^{(1,0)}+\lambda _1 \mu_{{N}_{2}{N}_{2}})}{\omega _{g1} \omega _{g2} \bar{N}_{2}}
\end{split}
\end{equation}

\subsection{The one-group one-point model for total detection of neutrons and gammas}
The motivation behind a separate derivation of the Feynman-alpha formula for the total detection of neutrons and gammas is related to the fact that variance of the total detection of neutrons and gammas cannot be represented as a linear combination of variances of the separate detection of neutrons and gammas because number of detections for neutrons and gamma are not independent variables. 

The assumptions below for the one-group one-point Feynman-alpha theory for the total detection of neutrons and gammas are similar to the ones used above for the separate detection of neutrons and gammas with the only difference that now ${Z(t,t_{d})}$ represents the number of total neutron and gamma detections in the time interval [${t_{d},t}$]. Thus, the joint probability of having $N_{1}$ neutrons and $N_{2}$ gammas at time $t$, and $Z$ neutrons and gammas together having been detected during the period of time ${t-t_{d}\geq 0}$ can be defined as $P(N_{1},N_{2},Z,t|t_0)$. Repeating the same procedure as before, one obtains the following forward-type equation:
\begin{equation}
\begin{split}
\frac{dP(N_{1},N_{2},Z,t)}{dt}\\
&=-(\lambda_{1}N_{1}+\lambda_{2}N_{2}+S)P(N_{1},N_{2},Z,t)\\
&+\lambda_{1a}(N_{1}+1)P(N_{1}+1,N_{2},Z,t)+\lambda_{2a}(N_{2}+1)P(N_{1},N_{2}+1,Z,t)\\
&+\lambda_{1f}\sum_k^{N_{1}+1}\sum_l^{N_{2}}(N_{1}+1-k)f(k,l)P(N_{1}+1-k,N_{2}-l,Z,t)\\
&+\lambda_{1d}(N_{1}+1)P(N_{1}+1,N_{2},Z-1,t)\\
&+\lambda_{2d}(N_{2}+1)P(N_{1},N_{2}+1,Z-1,t)\\
&+S\sum_m^{N_{1}}\sum_n^{N_{2}}p(m,n)P(N_{1}-m,N_{2}-n,Z,t)
\end{split}
\end{equation}
with the initial condition
\begin{equation}
\begin{split}
P(N_{1},N_{2},Z,t=t_0\mid t_0) = \delta_{N_{1},0} \delta_{N_{2},0} \delta_{Z,0} 
\end{split}
\end{equation}
A generating function with the initial condition for ${t_0 \leq t}$, ${G(X,Y,W,t=t_0\mid t_0) = 1}$, in this case will be introduced as below:
\begin{equation}
\begin{split}
G(X,Y,W,t)= \sum_{N_{1}} \sum_{N_{2}} \sum_{Z} X^{N_{1}} Y^{N_{2}} W^{Z}P(N_{1},N_{2},Z,t)
\end{split}
\end{equation}
After some simple manipulations, a partial differential equation is obtained:
\begin{equation}
\begin{split}
\frac{\partial G}{\partial t}= [\lambda_{1a} +\lambda_{1d}W - \lambda_{1}X + q(X,Y) \lambda_{1f}]\frac{\partial G}{\partial X} + [\lambda_{2a} +\lambda_{2d}W - \lambda_{2}Y]\frac{\partial G}{\partial Y} + S [r(X,Y) - 1] G
\end{split}
\end{equation}
For the stationary case, the quantities ${\bar{N}_{1}}$, ${\bar{N}_{2}}$, ${\bar{Z}}$ are given as follows:
\begin{equation}
\begin{split}
\bar{N}_{1} = \frac{S r^{(1,0)}}{\lambda _1-\lambda _{1f} q^{(1,0)}} \\
\bar{N}_{2} = \frac{S r^{(0,1)}}{\lambda _2}+\frac{S\lambda _{1f} q^{(0,1)}r^{(1,0)}}{\lambda _2(\lambda _1-\lambda _{1f} q^{(1,0)})} \\
\bar{Z}=(\lambda_{1d} \bar{N}_{1} +\lambda_{2d} \bar{N}_{2}) t\\
\end{split},
\end{equation}
where, for obvious reasons, the first two expressions are identical with those of the previous case, Eqs (\ref{eq:10}).
The modified second moments, ${\mu _{{{N}_{1}{N}_{1}}}}$, ${\mu _{{{N}_{1}{N}_{2}}}}$, ${\mu _{{{N}_{2}{N}_{2}}}}$ can be found by solving the system of equations in the stationary state with zero left hand sides:
\begin{equation}
\begin{split}
\frac{\partial}{\partial t} \mu_{{N}_{1}{N}_{1}} = 2(\lambda _{1f} q^{(1,0)}-\lambda _1) \mu_{{N}_{1}{N}_{1}}+\lambda _{1f} q^{(2,0)} \bar{Z}+S r^{(2,0)} \\
\frac{\partial}{\partial t} \mu_{{N}_{1}{N}_{2}} = S r^{(1,1)}+\lambda _{1f} q^{(1,1)} \bar{Z}-\lambda _{2}\mu_{{N}_{1}{N}_{2}} +(\lambda _{1f} q^{(1,0)}-\lambda _1) \mu_{{N}_{1}{N}_{2}}+\lambda _{1f} q^{(0,1)}\mu_{{N}_{2}{N}_{2}}\\
\frac{\partial}{\partial t} \mu_{{N}_{2}{N}_{2}} = S r^{(0,2)}-2\lambda _{2}\mu_{{N}_{2}{N}_{2}} +\lambda _{1f} q^{(0,2)}\bar{Z}+2\lambda _{1f} q^{(0,1)}\mu_{{N}_{1}{N}_{2}}
\end{split}
\end{equation}
The modified second moments, ${\mu _{{{N}_{1}{Z}}}}$, ${\mu _{{{N}_{2}{Z}}}}$, ${\mu _{{ZZ}}}$ can be found by solving the system of equations:
\begin{equation}
\begin{split}
\frac{\partial}{\partial t} \mu_{{N}_{1}{Z}} = \mu_{{N}_{1}{Z}} (\lambda _{1f} q^{(1,0)}-\lambda _1)+\lambda _{1d} \mu _{{N}_{1}{N}_{1}}+\lambda _{2d} \mu _{{N}_{1}{N}_{2}} \\
\frac{\partial}{\partial t} \mu_{{N}_{2}{Z}} =-\lambda _2 \mu_{{N}_{2}{Z}}+\lambda _{1f} q^{(0,1)} \mu_{{N}_{1}{Z}}+\lambda _{1d} \mu _{{N}_{1}{N}_{2}}+\lambda _{2d} \mu _{{N}_{2}{N}_{2}}\\
\frac{\partial}{\partial t} \mu_{ZZ} = 2 \lambda _{1d} \mu_{{N}_{1}{Z}} +2 \lambda _{2d} \mu_{{N}_{2}{Z}}
\end{split}
\end{equation}
Thus, the final expression of the Feynman-alpha formula for the total joint detection of neutrons and gammas is given as below:
\begin{equation}
\begin{split}
\frac{\sigma_{{ZZ}}^2(t)}{\bar{Z}} = 1 + Y_{t1} (1 - \frac{1-e^{- \omega_{t1} t}}{\omega_{t1} t}) + Y_{t2} (1 - \frac{1-e^{- \omega_{t2} t}}{\omega_{t2} t})
\end{split}
\end{equation}
where the functions ${Y_{t1}}$, ${Y_{t2}}$ are given in the form:
\begin{eqnarray}
& &-Y_{t1} = \frac{2 \lambda _{2d} (-\lambda _{1d} (\mu _{{N}_{1}{N}_{2}} (\lambda _{1f} q^{(1,0)}+\omega _{t1})-\lambda _{1f} \mu _{{N}_{1}{N}_{1}} q^{(0,1)}))}{\omega _{t1} (\omega _{t1}-\omega _{t2}) (\bar{N}_{1} \lambda _{1d}+\bar{N}_{2} \lambda _{2d})}\nonumber  \\
& + & \frac{2 \lambda _{2d} (-\lambda _{2d}(\mu _{{N}_{2}{N}_{2}} (\lambda _{1f} q^{(1,0)}+\omega _{t1})-\lambda _{1f} \mu _{{N}_{1}{N}_{2}} q^{(0,1)})+\lambda _1 (\lambda _{1d} \mu _{{N}_{1}{N}_{2}}+\lambda _{2d} \mu _{{N}_{2}{N}_{2}}))}{\omega _{t1} (\omega _{t1}-\omega _{t2}) (\bar{N}_{1} \lambda _{1d}+\bar{N}_{2} \lambda _{2d})} \nonumber \\
\end{eqnarray}
\begin{eqnarray}
& &-Y_{t2}=\frac{2 \lambda _{2d} (-\lambda _{1d} \mu _{{N}_{1}{N}_{2}} (\lambda _{1f} q^{(1,0)}+\omega _{t2})-\lambda _{1f} \mu _{{N}_{1}{N}_{1}} q^{(0,1)})}{\omega _{t2} (\omega _{t2}-\omega _{t1}) (\bar{N}_{1} \lambda _{1d}+\bar{N}_{2} \lambda _{2d})}\nonumber  \\
& + & \frac{2 \lambda _{2d} (-\lambda _{2d} (\mu _{{N}_{2}{N}_{2}} (\lambda _{1f} q^{(1,0)}+\omega _{t2})-\lambda _{1f} \mu _{{N}_{1}{N}_{2}} q^{(0,1)})+\lambda _1 (\lambda _{1d} \mu _{{N}_{1}{N}_{2}}+\lambda _{2d} \mu _{{N}_{2}{N}_{2}}))}{\omega _{t2} (\omega _{t2}-\omega _{t1}) (\bar{N}_{1} \lambda _{1d}+\bar{N}_{2} \lambda _{2d})} \nonumber \\
\end{eqnarray}
It can be shown that
\begin{eqnarray}
& &Y_{t0} = Y_{t1}+Y_{t2}\nonumber  \\
& = - & \frac{2 \lambda _{2d} (\lambda _{2d} \lambda _{1f} \mu _{{N}_{1}{N}_{2}} q^{(0,1)}-\lambda _{2d} \lambda _{1f} \mu _{{N}_{2}{N}_{2}} q^{(1,0)}+\lambda _1 \lambda _{1d} \mu _{{N}_{1}{N}_{2}}+\lambda _1 \lambda _{2d} \mu _{{N}_{2}{N}_{2}})}{\omega _{t1} \omega _{t2} (\bar{N}_{1}\lambda _{1d}+\bar{N}_{2} \lambda _{2d})} \nonumber  \\
& - & \frac{2 \lambda _{2d} (\lambda _{1d} \lambda _{1f} \mu _{{N}_{1}{N}_{1}} q^{(0,1)}-\lambda _{1d} \lambda _{1f} \mu _{{N}_{1}{N}_{2}} q^{(1,0)})}{\omega _{t1} \omega _{t2} (\bar{N}_{1}\lambda _{1d}+\bar{N}_{2} \lambda _{2d})}
\end{eqnarray}
It is interesting to notice that ${\omega_{t1}}$ and ${\omega_{t2}}$ are the same as in the case of variance to mean formula for gamma detections only:
\begin{equation}
\begin{split}
\omega_{t1}=-\lambda _{1f} q^{(1,0)}+\lambda _1\\
\omega_{t2}=\lambda _2
\end{split}
\end{equation}

\section{Analysis of the theoretical expressions for asymptotic values of the Feynman-Y functions for neutron, gamma and total detections (specific case)}
\subsection{A tendency in the behavior of the asymptotic values of the Feynman-Y function for the gamma, neutron and total detections}
Because of their use in safeguards applications, in order to see the tendency in the behavior of asymptotic values of the Feynman-Y function for the gamma, neutron and total detections, in this section we analyze the theoretical expressions for thee asymptotic values of the Feynman-Y functions for the specific case when only a compound source is present in a system, but no internal multiplication takes place. This means that one sets $\lambda_{1f} = 0$, which leads to a significant simplification of the formulas. All terms related to the first and second moments of the number of neutrons in induced fission will disappear. It is clear that these formulas could have been obtained in a simpler way, too. 

After a number of algebraic operations, the asymptotic values of the Feynman-Y function for the neutron, gamma and total detections can be expressed in a form as below:

\begin{eqnarray}\label{eq:33}
Y_{n,\infty} =\frac{\lambda_{1d}r^{(2,0)}}{\lambda_{1}r^{(1,0)}} \\
Y_{g,\infty} =\frac{\lambda_{2d}r^{(0,2)}}{\lambda_{2}r^{(0,1)}} \\
Y_{t,\infty} = Y_{g,\infty} \frac{(1+\frac{Y_{n,\infty}}{Y_{g,\infty}}\frac{2\lambda_{1}}{(\lambda_{1}+\lambda_{2})}\frac{(r^{(1,0)})^{2}}{r^{(2,0)}})}{(1+\frac{Y_{n,\infty}}{Y_{g,\infty}}\frac{(r^{(1,0)})^{2}}{(r^{(0,1)})^{2}}\frac{r^{(0,2)}}{r^{(2,0)}})} 
\end{eqnarray}
Thus, as it could be easily obtained from the expressions above, the asymptotic value of the Feynman-Y function for the total detections will be higher, lower or equal to that for gamma detections when following inequalities are valid :

\begin{eqnarray}\label{eq:35}
\frac{\lambda_2}{\lambda_1}<\frac{2(r^{(0,1)})^2}{r^{(0,2)}}-1\Rightarrow Y_{t,\infty} > Y_{g,\infty} \\
\frac{\lambda_2}{\lambda_1}>\frac{2(r^{(0,1)})^2}{r^{(0,2)}}-1\Rightarrow Y_{t,\infty} < Y_{g,\infty} \\
\frac{\lambda_2}{\lambda_1}=\frac{2(r^{(0,1)})^2}{r^{(0,2)}}-1\Rightarrow Y_{t,\infty} = Y_{g,\infty} 
\end{eqnarray}
In the case of a $^{252}$Cf-source the value of the expression $(\frac{2(r^{(0,1)})^2}{r^{(0,2)}}-1)$ is equal to 0.924959. 

Thus, one may conclude that the tendency in the behavior of the asymptotic values of the Feynman-Y function for the gamma, neutron and total detections will be different, depending on the properties of the system. 

It should be also mentioned that if we consider a two-group theory for neutrons \cite{Dinanew} for this specific case, the final expressions for an asymptotic value of the Feynman-Y function for neutron detection in non-multiplying media will be the same as Eq. (\ref{eq:33}) with the only difference that: 
\begin{eqnarray}
\lambda_1 = \lambda_{1a} + \lambda_{R} + \lambda_{1d},
\end{eqnarray}
where $\lambda_{R}$ is a removal of neutrons from a fast group (highly energetic neutrons) to a thermal group (low energetic neutrons).
 
\subsection{Evaluation of the source intensity via asymptotic values of the Feynman-Y function and count rate for the gamma, neutron and total detections}
It is interesting to notice that in a non-multiplying medium with a multiple (compound) source, or, better to say in the case of a small sample with negligible internal multiplication, one can estimate the source intensity (hence the sample mass) by using the asymptotic value of the Feynman-Y function and the average count rate (counts per second, cps) in the long gate (when an asymptotic value is achieved) for neutron and gamma detections as below: 
\begin{eqnarray}
\frac{Y_{n,\infty} }{cps_n}=\frac{r^{(2,0)}}{(r^{(1,0)})^2}\frac{1}{S} \\
\frac{Y_{g,\infty} }{cps_g}=\frac{r^{(0,2)}}{(r^{(0,1)})^2}\frac{1}{S}  
\end{eqnarray}

However, as could be seen from the expression below, it is not easy to estimate it by using an asymptotic value of the Feynman-Y function for total detections:
\begin{eqnarray}
\frac{Y_{t,\infty}cps_t }{(Y_{g,\infty})^2}=S\frac{(r^{(0,1)})^2}{r^{(0,2)}}(1+\frac{2(r^{(0,1)})^2}{r^{(0,2)}}\frac{(\frac{cps_t}{cps_g}-1)}{(1+\frac{\lambda_2}{\lambda_1})})
\end{eqnarray}
where as follows from $\frac{\lambda_2}{\lambda_1}>0$:
\begin{eqnarray}
\frac{2(r^{(0,1)})^2}{r^{(0,2)}}>1
\end{eqnarray}
Nevertheless, inclusion of the information from the asymptotic value of the Feynman-Y function for the total detection of neutrons and photons might improve the accuracy of the evaluation of the source intensity.

\section{Numerical illustration of the new versions of the Feynman-alpha theory and their comparison to the traditional variance to mean ratio for neutrons}
\begin{figure}[ht!]
\centering
\includegraphics[width=0.95\textwidth]{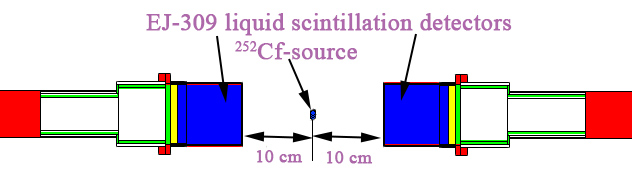}
\caption{Geometry used for the Monte-Carlo simulations.}
\label{fig:3}
\end{figure}

In order to compare quantitatively the three different alternatives of the Feynman-alpha theory, namely the traditional one and the two new ones introduced in this paper, MCNPX \cite{MCNPX} simulations were performed in a simplified setup. The setup consists of two EJ-309 liquid scintillation detectors (D76 x 76 mm) and a $^{252}$Cf-source in a form of a disk with a radius of 0.45 cm, as shown in Figure \ref{fig:3}. Each detector is located 10 cm from the source.

This setup was chosen because of its similarity to a potential measurement situation with an orphan source. As mentioned before, the conditions needed for the validity of the formulas derived in the foregoing are not fulfilled. However, as was also mentioned, only the asymmpotic values of the variance to mean will be used, which are supposed to be less affected by the deviations between the measurement conditions and the assumptions of the theory. 

The comparison of the neutron, gamma and total (neutron and gamma) variance to mean ratios is made by using quantitative values of the transition probabilities and reaction intensities obtained in a way similar to that described in \cite{Dinanew,ADSFreya,Chernikova2013,Anderson20121,Anderson2012,Chernikova20131}. In addition, some quantities were obtained via processing a ptrac file. Gammas from neutron capture are not included either in the simulations or in the theory; they will be considered in further work. The coefficients obtained in MCNPX simulations and used for building up theoretical variance to mean ratios are shown in Table 1 and can be easily modified (if needed) in an interactive Mathematica notebook for visualization of gamma and total (neutron-gamma) Feynman-Alpha formulas\footnote{The interactive Mathematica notebook can be downloaded from dx.doi.org/10.13140/2.1.1395.1684}. A detection efficiency ($\varepsilon$) for neutrons and gammas is assumed to be equal to 1\%, $\lambda_{d}=\varepsilon \cdot \lambda_{a}$ in which the geometric efficiency is inteded to be included.  

\begin{table}[htbp]\footnotesize
\caption{The quantitative values of coefficients, the transition probabilities and reaction intensities.}
  \centering
    \begin{tabular}{rrrr}
\\    
    \hline
    \hline
          & \textbf{$\lambda_{a} ({\scriptsize \dfrac{1}{s}})$} & \textbf{$\lambda_{f} ({\scriptsize \dfrac{1}{s}})$ } & \textbf{$\lambda_{d} ({\scriptsize \dfrac{1}{s}})$} \\
          \hline
          \hline
    $neutrons$    & 0.00347 & 0 & 0.0000347   \\
    $gammas$   &0.0093161 & 0 & 0.000093161 \\
    \\
    \hline
    \hline
     \textbf{$r^{(1,0)}$}  & \textbf{$r^{(0,1)}$} & \textbf{$r^{(2,0)}$} & \textbf{$r^{(0,2)}$} \\
     \hline
     \hline
    3.7639 & 7.9948 & 11.9718 & 66.4085   \\
    \end{tabular}%
\label{table:tab2}
\end{table}%

As one can see, $\frac{\lambda_2}{\lambda_1}=2.68476$ which, according to Eq. (\ref{eq:35}) indicates that ${Y_{t,\infty} < Y_{g,\infty}}$ and ${Y_{g,\infty} > Y_{n,\infty}}$.

\begin{figure}[ht!]
\centering
\includegraphics[width=0.85\textwidth]{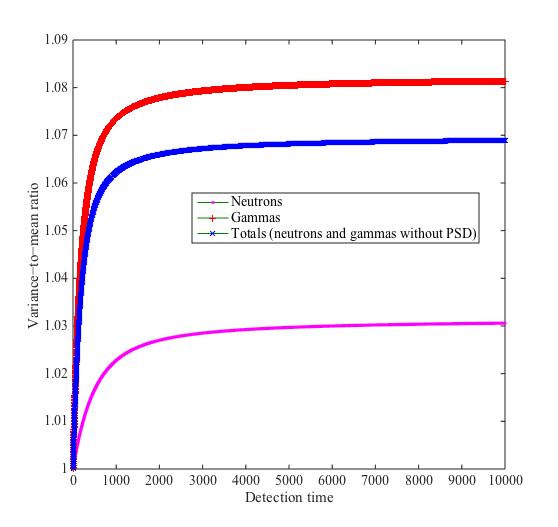}
\caption{The dependence of the variance to mean of the number of neutron, gamma and total (neutron and gamma) detections on the detection time.}
\label{fig:4}
\end{figure}

As shown in Figure \ref{fig:4}, the behaviour of the dependence of the variance to mean for the number of gamma, neutron and total (neutron and gamma) detector counts on the detection time agrees with the theory predictions. The asymptotic value of the Feynman-Y function is higher for gamma detections in comparison with the total and neutron detections.

The variance to mean ratio for neutrons\footnote{In the case of using EJ-309 liquid scintillation detectors we only consider neutrons with light output higher than 200 keVee} reaches its asymptotic value in a time range between 500-1000 ns, whereas the variance to mean ratio for gamma detections and total detections result in a plateau starting in the time interval of 1-500 ns. Formally, this is due to the fact that in addition to the single exponent of the neutron variance to mean formula, the gamma and neutron-gamma variance to mean contains one more exponent, which is larger, due to the shorter lifetime of the gamma photons.

\section{Experimental illustration of the new versions of the Feynman-Y theory and their comparison to the traditional variance to mean ratio for neutrons}
\subsection{Description of an experimental set-up and procedure}
In addition to the numerical evaluation of the newly derived variance to mean formulas for gamma and total (neutron and gamma) detections, these ratios were evaluated experimentally. Because of the earlier mentioned deviations between the assumptions of the theory and the experimental setup, these should be considered more like an attempt to get hands-on experience with the data collection and and evaluation procedure, as well as a preliminary study of the effect of the approximations of the theory. 

\begin{figure}[ht!]
\centering
\includegraphics[width=0.85\textwidth]{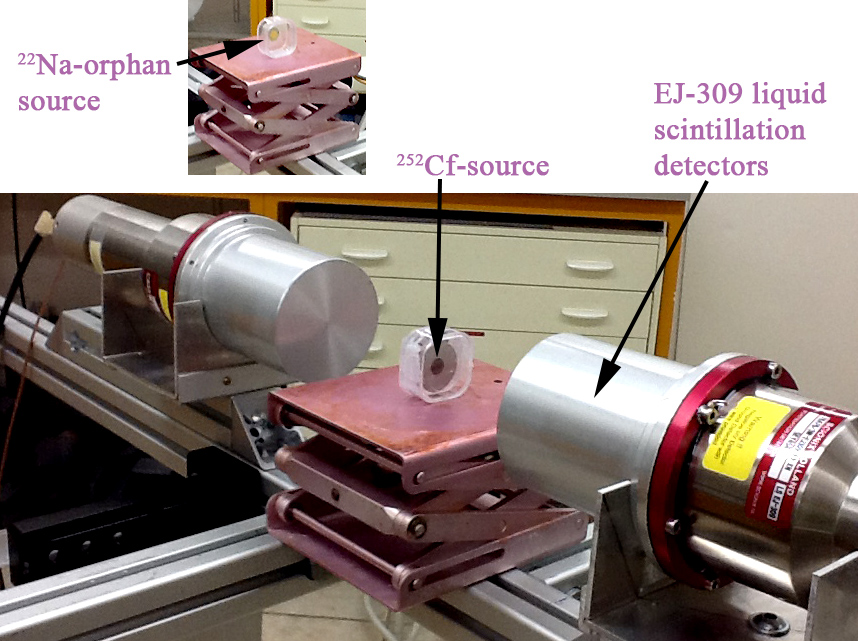}
\caption{Photo of the configuration of the experimental setup.}
\label{fig:5}
\end{figure}

The experiments were performed with a weak $^{252}$Cf (${\sim}$17.275 kBq) neutron-gamma source (originally, $^{252}$Cf ionization chamber detector), a $^{137}$Cs random gamma source (${\sim}$22.498 kBq), $^{22}$Na correlated gamma source (${\sim}$ 2125 kBq) and an orphan $^{22}$Na (activity is known to be less than 1 MBq). The $^{252}$Cf-source is a disk with a radius of 0.45 cm. The experimental setup consisted of two EJ-309 (D76 x 76 mm) liquid scintillation detectors, each located at a distance of 10 cm from the source and one of the two sources, i.e. $^{252}$Cf and $^{137}$Cs, as shown in Figure \ref{fig:5}. In the case of measurements with the $^{22}$Na orphan-source and a strong $^{22}$Na known source, the distances were increased up to 20 cm and 50 cm, respectively, to decrease the amount of pile-up events. The detectors were connected to a 8 channel, 12 bit 250 MS/s, VX1720E CAEN digitizer.

\begin{figure}[ht!]
\centering
\includegraphics[width=0.85\textwidth]{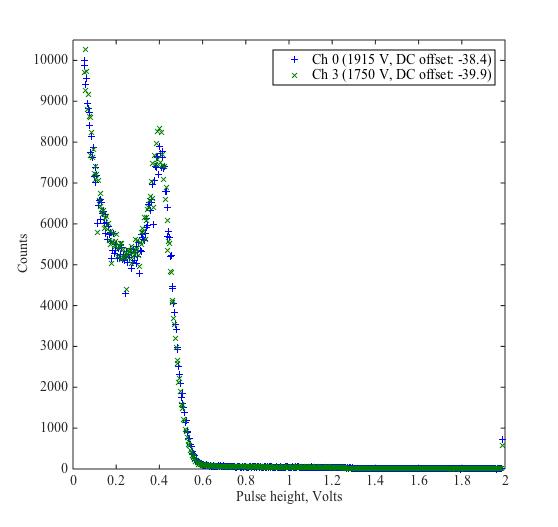}
\caption{Pulse height spectra obtained for two detectors/channels with a $^{137}$Cs source.}
\label{fig:6}
\end{figure}

As a first step of the experimental work, a calibration procedure was performed with a $^{137}$Cs source. Pulse height spectra were independently collected during 5 minutes for each detector with a non-overlapping trigger. The results of the calibration are shown in Figure \ref{fig:6}. As a result of this procedure, the high voltage bias and the DC offset were adjusted for each detector/channel individually, as follows: Channel 0 (Ch 0, Voltage: 1915 V, DC offset: -38.4), Channel 3 (Ch 3, Voltage: 1750 V, DC offset: -39.9).

\begin{figure}[ht!]
\centering
\includegraphics[width=0.85\textwidth]{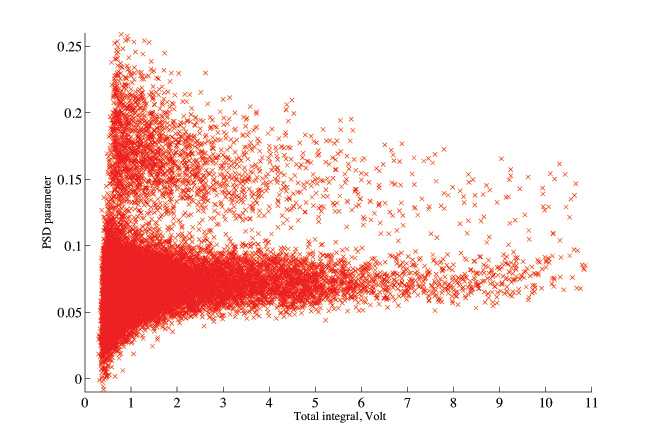}
\caption{2D plot of the counts as a function of the pulse energy and the PSD parameter\cite{Boron}.}
\label{fig:7}
\end{figure}

The experimental evaluation of the variance to mean ratios for neutron, gamma and total detections was done with two detectors. Data were collected during 200 seconds using a simultaneous trigger for all detectors. The data from the detectors were post processed offline to obtain the dependence of the ratio of the variance-to-mean of the number of neutron, gamma and total (neutron and gamma) detections on the detection time (with 0.2 V threshold). This was done the traditional reactor-physics way \cite{Imre,Kitamura} via evaluating the variance and mean of the numbers of counts (N) in k consecutive time intervals of length T, as shown below:
\begin{equation*}
\begin{split}
Variance_{k}(T)=\frac{1}{k-1}\cdot \sum_{i=1}^{k}(N_{i}-\frac{1}{k}\sum_{i=1}^{k}N_{i})^{2} \\
Mean_{k}(T)=\frac{1}{k}\sum_{i=1}^{k}N_{i}
\end{split}
\end{equation*}
The uncertainty in the variance to mean ratio was estimated as below \cite{Okowita}:
\begin{equation*}
\begin{split}
\sigma_{Y}^{2}[T]=\frac{(\overline{N^{2}})^{2}}{(\overline{N})^{2}}\cdot (\frac{(\overline{N^{4}}-\overline{N^{2}}^{2}))}{(\overline{N^{2}})^{2}\cdot (k-1))}+\frac{(\overline{N^{2}}-\overline{N}^{2}))}{(\overline{N})^{2}\cdot (k-1))}-\frac{2\cdot (\overline{N^{3}}-\overline{N^{2}}\cdot \overline{N}))}{(\overline{N^{2}}\cdot \overline{N})\cdot (k-1))})+\frac{(\overline{N^{2}}-\overline{N}^{2}))}{(k-1))},
\end{split}
\end{equation*}
In case of collecting data from a $^{252}$Cf-source, pulse shape discrimination was performed in a way similar to \cite{Boron} by using a charge comparison method \cite{Charge}. In Figure \ref{fig:7} it is seen that the neutrons and gammas are well separated.
Thus, for a sufficiently good separation between neutrons and gammas a discrimination can be achieved by just setting the PSD parameter threshold equal to 0.115. More accurate separation requires application of another method for neutron-gamma discrimination, e.g. correlation-based techniques \cite{102} or artificial neural networks \cite{101,103}.

\subsection{Experimental variance-to-mean ratios for neutron, gamma and total detections}

As a first step, using the $^{252}$Cf source, the source strength was extracted from the asymptotic value of the variance to mean for counting neutrons only. The source strength was correctly estimated from the measurements. Quantitative results are shown in Fig. \ref{fig:9}
below. Such estimates were performed with success also by other group \cite{Nakae}. This means that for the case of neutrons, the deviations between the model assumptions and the experiments do not play a significant role.

\begin{figure}[ht!]
\centering
\includegraphics[width=0.85\textwidth]{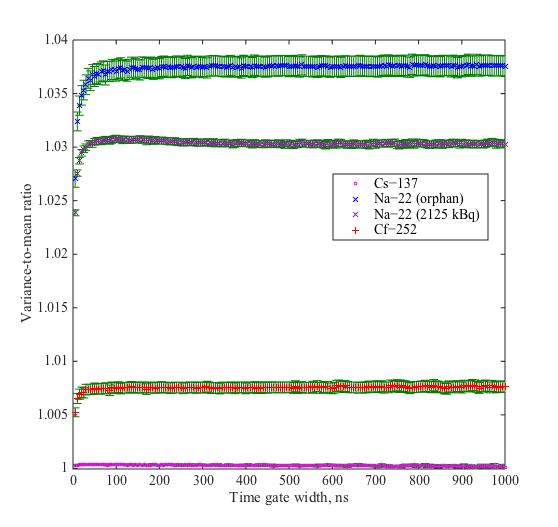}
\caption{The dependence of the ratio of the variance-to-mean of the number of gamma detections on the detection time for $^{252}$Cf, $^{137}$Cs and two $^{22}$Na sources.}
\label{fig:8}
\end{figure}

The test of the use of the newly elaborated formulas for gamma and gamma+neutron counts was performed in the second step. An experimental evaluation of the variance to mean ratios for gamma detections was performed for four sources: $^{252}$Cf, $^{137}$Cs, $^{22}$Na and $^{22}$Na-orphan. The results are shown in Figure \ref{fig:8}.

As was expected, in case of a Poisson gamma source, such as $^{137}$Cs, the value of the variance of the gamma detections is equal to the mean value. In contrast, $^{22}$Na is a non-Poisson (compound Poisson) source of gammas\footnote{A $^{22}$Na nucleus emits, in addition to a prompt gamma, also a positron, which due to annihilation leads to emission of two gammas.}, the same way as the $^{252}$Cf source emits multiple gammas. Therefore, the value of the variance is deviating from the mean value of the gamma detections for both sources. It is interesting to notice that the asymptotic values of the variance to mean ratio for two $^{22}$Na-sources are higher than for the $^{252}$Cf source, despite the fact that in general there are more gammas emitted in a $^{252}$Cf-source event in comparison to a $^{22}$Na-source event. However, the strength of the $^{22}$Na sources is at least an order of magnitude higher than the strength of $^{252}$Cf. 

An attempt was made to determine the possibility of evaluating the activity of an orphan $^{22}$Na source and a strong $^{22}$Na known-source using the asymptotic values of the Feynman-Y function and count rate for the gamma detections and equation 41.  
\begin{table}[ht!]\footnotesize
\caption{Evaluation of the activity of $^{22}$Na-sources via asymptotic values of the Feynman-Y function and count rate for the gamma detections.}
  \centering
    \begin{tabular}{rrrrr}
\\    
    \hline
    \hline
   Declared activity & Feynman-$Y_{g,\infty}$  & Count rate, cps & Solid angle & Calculated source activity \\
    &   &  & correction factor &  \\
     \hline
     \hline
 2125 kBq & 0.0303$\pm$0.0003 & 5957$\pm$4 & 346.26 & 46895$\pm$465 kBq \\
  orphan source      & 0.0376$\pm$0.0009 & 400.36$\pm$1.41 & 55.40 & 406.32$\pm$9.83 kBq
    \end{tabular}%
\label{table:tab3}
\end{table}%
The results of this evaluation are shown in Table 2. Unfortunately, as it is shown for a $^{22}$Na source with known activity, the calculated value is 22 times higher than the declared value. This indicates that for the case using gamma detections, the deviations between the model assumptions and the experiment play a larger role than for neutrons. Some of the deviations may be sought in the experimental settings, and in particular in the detection process (e.g. the type of detectors used), since in a liquid scintillation detector each individual photon undergoes a multiple number of Compton scatterings etc., while in a theory it is assumed that the first detection removes particles from the system.

One, of course, may try to assume that the ratio between the declared activity and the calculated activity should be the same for both $^{22}$Na-sources. Such a procedure would yield the $^{22}$Na orphan source  has an activity of 18.3975 kBq. However, this is not a reliable approach. Therefore, to solve this issue, in further work we plan to perform similar measurements with HPGe detectors.      

Since the $^{252}$Cf is a source of both neutrons and gammas, the next step of the investigation was the evaluation of the variance to mean ratio for neutron, gamma and total\footnote{No neutron-gamma discrimination was applied to obtain the total counts} detections. 
\begin{figure}[ht!]
\centering
\includegraphics[width=0.85\textwidth]{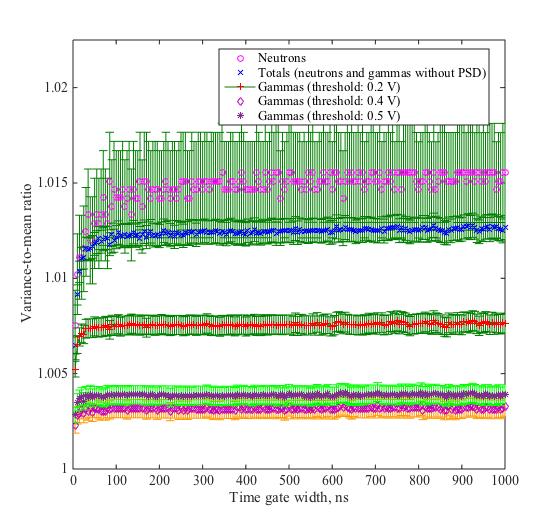}
\caption{The dependence of the ratio of the variance to mean of the number of neutron, gamma (thresholds: 0.2 V, 0.4 V, 0.5 V) and total detections on the detection time for $^{252}$Cf-source.}
\label{fig:9}
\end{figure}
As shown in Figure \ref{fig:9}, the asymptotic value of variance to mean ratio for neutron detections is higher than that for gammas (for all thresholds, i.e. 0.2 V, 0.4 V, 0.5 V) and totals. At the same time, in contrast to the theoretical predictions, the asymptotic value of variance to mean ratio for total and neutron detections is overestimated. 

\begin{table}[ht!]\footnotesize
\caption{Evaluation of the $^{252}$Cf-source activity via asymptotic values of the Feynman-Y function and count rate for the gamma and neutron detections.}
  \centering
    \begin{tabular}{rrrrr}
\\    
    \hline
    \hline
   Feynman-$ Y_{i,\infty}$ & $Y_{n,\infty}$  & $Y_{g,\infty}$, 0.2 V & $Y_{g,\infty}$, 0.4 V & $Y_{g,\infty}$, 0.5 V \\
     \hline
     \hline
    & 0.0147$\pm$0.0025 & 0.0075$\pm$0.0005 & 0.0031$\pm$0.0005 & 0.0039$\pm$0.0005   \\
    \\
    \hline
    \hline
 Count rate, cps   & $cps_n$  & $cps_g$, 0.2 V & $cps_g$, 0.4 V & $cps_g$, 0.5 V\\
     \hline
     \hline
    & 22.51$\pm$0.36 & 301.55$\pm$1.23 & 168.95$\pm$0.92 & 135.38$\pm$0.82   \\
        \\
    \hline
    \hline
 Calculated source activity, Bq   & $I_n$  & $I_g$, 0.2 V & $I_g$, 0.4 V & $I_g$, 0.5 V\\
     \hline
     \hline
    & 17923$\pm$3061 & 578590$\pm$38645 & 784270$\pm$12657 & 499530$\pm$64114 \\
 \textbf{Declared source activity}& \textbf{17275 Bq} &    &  &  
    \end{tabular}%
\label{table:tab4}
\end{table}%
It should be also mentioned that for the case of gamma and total detections, a measurement period of 200 seconds was sufficient to get the results with relatively low uncertainty, while for the case of neutron detections, the uncertainties are significant for comparable measurement times.

As it is shown in Table 3, the activity of the $^{252}$Cf source, calculated from the asymptotic values of the Feynman-Y function and count rate for the gamma detections is 22-33 times higher than the declared value for all thresholds used. As mentioned earlier, the most likely reason is the deviation between the model assumptions and experimental setup, including the detection process, just as with the  $^{22}$Na sources. This is in contrast to the fact that, as also mentioned earlier, the activity of $^{252}$Cf calculated via using asymptotic values of the Feynman-Y function and count rate for the neutron detections is in agreement with the declared value. The reasons for the discrepancy for the gamma and neutron-gamma detections will be investigated both theoretically and experimentally in future work.

\section{Conclusion}
Derivation and evaluation of analytical expressions for the new neutron-gamma variance to mean (Feynman-alpha) formulas for separate gamma and total detections for both reactor and safeguards applications were presented in this paper. It was found that the variance to mean (Feynman-alpha) formulas for separate gamma and total detections are both obtained in a two-exponential form. One of the exponents (${\omega_1}$) is the same for total and gamma detections as for neutron detections. This lead to the conclusion that the variance to mean (Feynman-alpha) formulas for separate gamma and total detections contain the same information on neutron population and neutron detection characteristics. This is not valid in reverse, i.e. the variance to mean (Feynman-Y) formula for neutron detections does not contain any information on gamma population or detection characteristics. 

The results of the analysis of the theoretical expressions for a case of non-multiplying media (source without internal multiplication) showed that the order of magnitude of asymptotic values for  
a variance to mean ratio for neutron, gamma and total detections depends on the properties ($\frac{\lambda_2}{\lambda_1}$) of the system and inherent properties of the source ($(\frac{2(r^{(0,1)})^2}{r^{(0,2)}}-1)$). In addition, the theoretical expressions derived for evaluation of the source intensity based on asymptotic values for a variance to mean ratio and count rate for the gammas, neutrons and totals showed that totals solely can not be used for this purpose due to the strong dependence of a final formula on external parameters ($cps_g$, $Y_{g,\infty}$, $\frac{\lambda_2}{\lambda_1}$). However, they might be used for improving an accuracy of already evaluated values.

As a whole, numerical evaluation of the theoretical formulas showed an agreement with theoretical predictions, although did not show an agreement with following experimental investigations for total and gamma detection. Most likely the reason for this is due to the deviations between the model assumptions underlying the theory and experiments, in particular the difference between the detection process in the experimental setup and the way how the detection process is described by theory. As a possible solution one may use different type of detectors, e.g. HPGe detector or CdZnTe. Interesting to notice is that the asymptotic values of the Feynman-Y function and count rate for the fast neutron detections and use of the equation 40 give a good prediction of the source activity.  

It is also worth mentioning that for the case of gamma and total detection time of 100 seconds for measurements were sufficient to obtain results with relatively low uncertainty, while for the case of neutron detections much larger uncertainties prevail for comparable measurement times.

Thus, one may conclude that the new formulas for gamma and total neutron-gamma detections have a promise to complement, and in some cases replace the traditional variance to mean (Feynman-Y) formula for neutron detections. The variance to mean (Feynman-Y) formula for total neutron-gamma detections may eliminate the problem related to discrimination between neutron and gamma particles in scintillation detectors. At the same time the variance to mean (Feynman-alpha) formula for gamma detections has a potential to be used with detectors of only gamma radiation, which is normally employed in a spent fuel pool.

\section*{Acknowledgement}
This work was supported by the Swedish Radiation Safety Authority, SSM, and the FP7 EU Collaborative Research Project FREYA, Grant Agreement no. FP7-269665. The authors want to thank \textbf{Dr. Lars Hildingsson} for useful discussions and advice.


\begin{thebibliography}{14}
\bibitem{Freya}{A. Kochetkov, e. al., Current progress and future plans of the FREYA Project,  Proceedings of Int. Conference, Technology and Components of Accelerator Driven Systems (TSADS-2), Nantes, France, 2013.}

\bibitem{Twomey} {T.R. Twomey, R.M. Keyser, Characteristics and Performance of an Integrated Portable High Efficiency Neutron Multiplicity Counter for Detection of Illicit Neutron Sources, in: S.I.A. ORTEC, Oak Ridge, TN, 37831 USA.}

\bibitem{Verbeke1} {J.M. Verbeke, G. Chapline, L. Nakae, et.al, Monitoring spent or reprocessed nuclear fuel using fast neutrons,  LLNL-CONF-490184.}

\bibitem{croft12} S.Croft, A.Favalli, D.Hauck, D.Henzlova, P.Santi,  Feynman Variance-to-mean in the context of passive neutron coincidence counting, Nuclear Instruments and Methods, Vol.686, 2012, pp. 136-144

\bibitem{McConchie} {S. McConchie, P. Hausladen, J. Mihalczo, Prompt Neutron decay Constant from Feynman Variance fitting, Oak Ridge National Laboratory; Bethel Valley Road, Oak Ridge, TN (Ed.), 01/2009.}

\bibitem{Verbeke} {J.M. Verbeke, L. Nakae, P. Kerr, et.al, Testing of liquid scintillator materials for gamma and neutron detection,  INMM, Tucson, USA, 2009.}
\bibitem{Nakae} {L. Nakae, et.al, Recent developments in neutron detection and multiplicity counting with liquid scintillator,  2nd Japan IAEA Workshop on Advanced Safeguards Technology for Future Nuclear Fuel Cycle, Tokai, Japan, 2009.}

\bibitem{MC2015} {Dina Chernikova, Imre P\'{a}zsit, Stephen Croft, Andrea Favalli, An effect of capture gammas, photofission and photonuclear neutrons to the neutron-gamma Feynman variance-to-mean ratios (neutron, gamma and total), submitted to ANS MC2015, Nashville, Tennessee, April 19, 2015.}

\bibitem{Dinanew} {D. Chernikova, W. Ziguan, I. P\'{a}zsit, L. P\'{a}l, A general analytical solution for the variance-to-mean Feynman-alpha formulas for a two-group two-point, a two-group one-point and a one-group two-point cases, The European Physical Journal Plus, 129 (2014) 259.}

\bibitem{Imre} {I. P\'{a}zsit, L. P\'{a}l, Neutron Fluctuations: A Treatise on the Physics of Branching Processes, Elsevier Science Ltd., London, New York, Tokyo, 2008.}

\bibitem{ads1} P\'{a}zsit I. and Yamane Y. Theory of neutron fluctuations in source-driven subcritical systems. Nucl. Instr. Meth. A 403 (1998) 431 - 441. 

\bibitem{MCNPX}{D. B. Pelowitz, MCNPX User’s Manual, Version 2.7.0, Los Alamos National Laboratory report LA-CP-11-00438, April 2011.}
\bibitem{ADSFreya}{P. Cartemo, A. Nordlund, D. Chernikova, Sensitivity of the neutronic design of an Accelerator-Driven System (ADS) to the anisotropy of yield of the neutron generator and variation of nuclear data libraries,  ESARDA meeting, Brugge, 2013.}

\bibitem{Chernikova2013} {D. Chernikova, I. P\'{a}zsit, W. Ziguan, Application of the two-group - one-region and two-region one-group Feynman-alpha formulas in safeguards and accelerator-driven system (ADS).  ESARDA meeting, Brugge, 05/2013.}

\bibitem{Anderson20121}{J. Anderson, D. Chernikova, I. P\'{a}zsit, L. P\'{a}l, S.A. Pozzi, Two-point theory for the differential self-interrogation Feynman-alpha method, The European Physical Journal Plus, 127 (2012) 90.}

\bibitem{Anderson2012} {J. Anderson, L. P\'{a}l, I. P\'{a}zsit, D. Chernikova, S. Pozzi, Derivation and quantitative analysis of the differential self-interrogation Feynman-alpha method, The European Physical Journal Plus, 127 (2012) 21.}

\bibitem{Chernikova20131} {D. Chernikova, I. P\'{a}zsit, L. P\'{a}l, Z. Wang, Derivation of two-group two-region Feynman-alpha formulas and their application to Safeguards and accelerator-driven system (ADS).  INMM 54th Annual Meeting, JW Marriott Desert Springs, Palm Desert, California USA, 07/2013.}

\bibitem{Kitamura} {Y. Kitamura, T. Misawa, A. Yamamoto, Y. Yamane, C. Ichihara, H. Nakamura, Feynman-alpha experiment with stationary multiple emission sources, Progress in Nuclear Energy, 48 (2006) 569-577.}

\bibitem{Okowita} {A. Okowita, J. Mattingly, Analysis of the Feynman variance to mean ratio using nonlinear  regression,  INMM 53rd Annual Meeting, Orlando, Florida, USA, 2012.}

\bibitem{Boron} {D. Chernikova, K. Axell, I. P\'{a}zsit, A. Nordlund, R. Sarwar, A direct method for evaluating the concentration of boric acid in a fuel pool using scintillation detectors for joint-multiplicity measurements, Nuclear Instruments and Methods in Physics Research Section A: Accelerators, Spectrometers, Detectors and Associated Equipment, (11/2013) 90–97.}

\bibitem{Charge} {F.D. Brooks, A scintillation counter with neutron and gamma-ray discriminators, Nuclear Instruments and Methods, 4 (1959) 151–163.}

\bibitem{101}{T. Tambouratzis, D. Chernikova, I. P\'{a}zsit, Pulse shape discrimination of neutrons and gamma rays using Kohonen artificial neural networks, Journal of Artificial Intelligence and Soft Computing Research, Vol.3, No.1,pp.77-88, 2013.}

\bibitem{102}{D. Chernikova, Z. Elter, K. Axell, A. Nordlund, A method for discrimination of neutron and gamma pile-up events in scintillation detectors with a simultaneous identification of malfunctioning ones,  INMM 55th Annual Meeting, Atlanta, Georgia, USA, 2014.}

\bibitem{103}{T. Tambouratzis, D. Chernikova, I. P\'{a}zsit, A Comparison of Artificial Neural Network Performance: the Case of Neutron Gamma Pulse Shape Discrimination,  IEEE Symposium on Computational Intelligence for Security and Defense Applications (CISDA), Singapore, 2013.}
\end{thebibliography}
\end{document}